\documentclass{iaus260}
\usepackage{graphicx}
\pubyear{2009}
\volume{260}  
\pagerange{10--17}
\jname{The R\^ole of Astronomy in Society and Culture}
\editors{D. Valls-Gabaud \& A. Boksenberg, eds.}

\title[Astronomy in Mozambique]{Introducing Astronomy into \\
	Mozambican Society}   

\author[Ribeiro et al.]{V. A. R. M. Ribeiro,$^1$ C. M. Paulo,$^{2,3}$ A. M. A. R. Besteiro,$^4$ \\
	H. Geraldes,$^5$ A. M. Maphossa,$^3$ F. A. Nhanonbe$^6$ \\
	\and A. J. R. Uaissine$^7$}   

\affiliation{$^1$Astrophysics Research Institute, Liverpool John Moores University, Twelve Quays House, Egerton Wharf, Birkenhead, Wirral, CH41 1LD, UK \\ email: {\tt var@astro.livjm.ac.uk} \\ [\affilskip]
	$^2$Physics Department, University of the Western Cape, Modderdam Rd, Bellville, 7535, \\ Cape Town, South Africa \\ [\affilskip]
	$^3$Departamento de F\'isica, Universidade Eduardo Mondlane, Maputo, Mo\c{c}ambique \\ email: {\tt claudiompaulo@uem.mz, maphossa@zebra.uem.mz} \\ [\affilskip]
	$^4$Escola Portuguesa de Mo\c{c}ambique, Av. para o Palmar, 562, \\
	Caixa Postal 2490, Maputo, Mo\c{c}ambique \\ email: {\tt abesteiro@epmcelp.edu.mz}\\ [\affilskip]
	$^5$Amateur Astronomer \\ email: {\tt geraldes@gmail.com} \\ [\affilskip]
	$^6$Instituto Nacional de Meteorologia, Rua de Mukumbura, 164, \\
	Caixa Postal 256, Maputo, Mo\c{c}ambique \\ email: {\tt faustino.armando@yahoo.com.br} \\ [\affilskip]
	$^7$Minist\'erio da Ci\^encia e Tecnologia, Av. Patrice Lumumba, 770, \\ 
	Maputo Mo\c{c}ambique \\ email: {\tt antonio.uaissone@mct.gov.mz} }

\begin{document}
\maketitle

\begin{abstract} 
Mozambique has been proposed as a host for one of the future Square Kilometre Array stations in Southern Africa. However, Mozambique does not possess a university astronomy department and only recently has there been interest in developing one. South Africa has been funding students at the MSc and PhD level, as well as researchers. Additionally, Mozambicans with Physics degrees have been funded at the MSc level. With the advent of the International Year of Astronomy, there has been a very strong drive, from these students, to establish a successful astronomy department in Mozambique. The launch of the commemorations during the 2008 World Space Week was very successful and Mozambique is to be used to motivate similar African countries Ð who lack funds but are still trying to take part in the International Year of Astronomy. There hare been limited resources and funding, however there is a strong will to carry this momentum into 2009 and, with this, influence the Government to introduce Astronomy into its national curriculum and at University level. Mozambique's motto for the International Year of Astronomy is ``Descobre o teu Universo".
\end{abstract}

\firstsection

\section{Introduction}
\begin{figure}[h]
\centering
\includegraphics[width=3.4in]{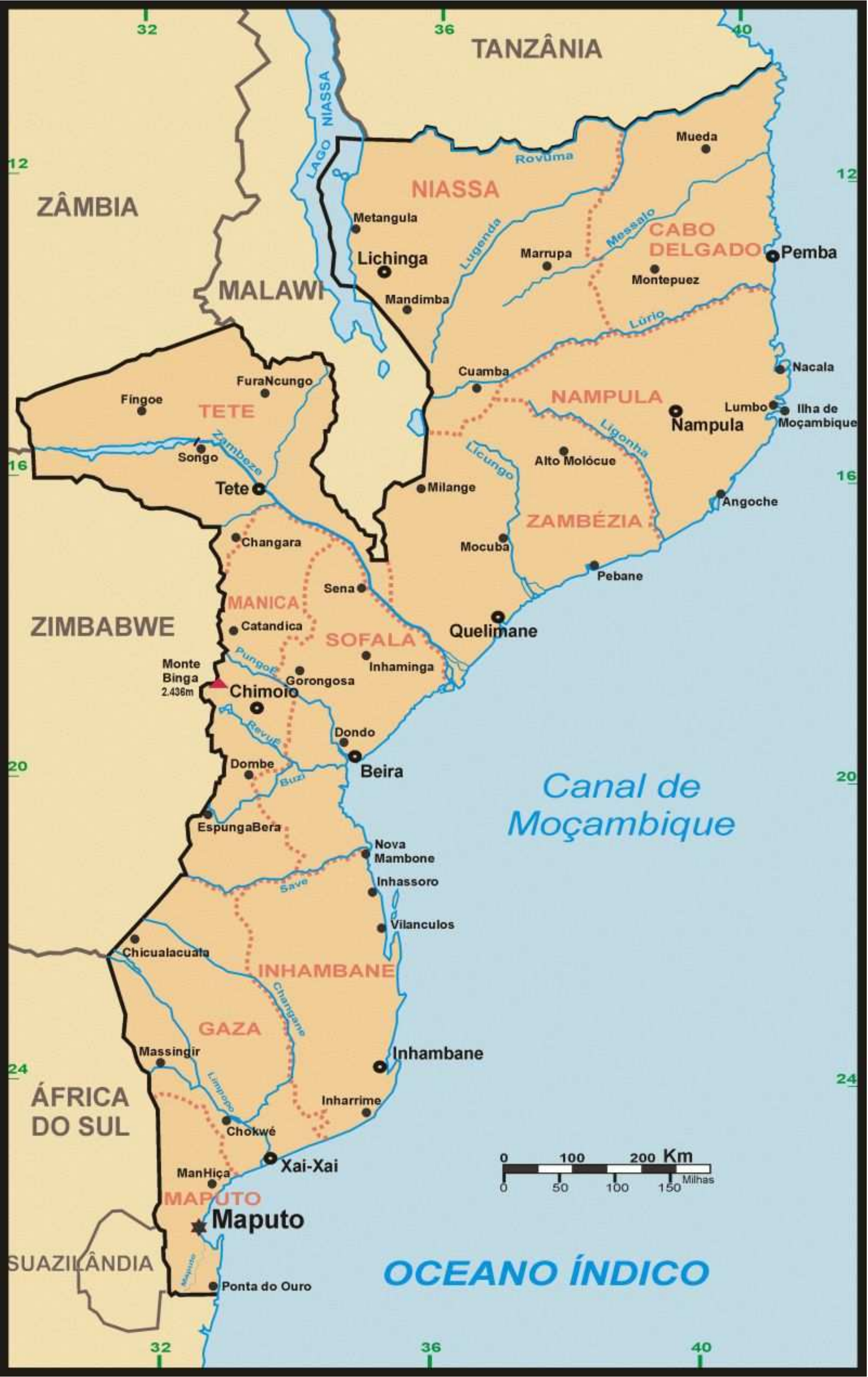}
\caption{Map of Mozambique showing its different cities and bordering countries. Source: http://www.portaldogoverno.gov.mz/Mozambique/mapa\_mocambique.jpg, accessed 2009 May 13.}
\label{fig:1}
\end{figure}

Mozambique is located in the south-eastern coast of Africa boarded by the Indian Ocean in the east, Tanzania to the north, Malawi and Zambia to the north-west, Zimbabwe to the west and South Africa and Swaziland to the south-west (Figure \ref{fig:1}). Mozambique's most recent social history comprises of a war of independence from the former Portuguese colonial rulers in 1975 June 25 and then followed two years later by a civil war until 1992 October 4. However, ever since Mozambique has enjoyed stability.

According to the 2007 population census, Mozambique is inhabited by just over 20 million people [\cite{ine}]. The official language in Mozambique is Portuguese, although it is not spoken by everyone, and a further 20 languages exist. The educational system is free and compulsory until the age of 18. There are milestones to cross in years 7, 10 and 12, in the form of standardised national exams to allow continuity into secondary school, pre-University years and eventually to finish school or to enter university, respectively. However, places at the universities are limited and a further test is required to allow access into these institutions.

There is a big bottle-neck in the educational system where you have millions of young people in primary school and then only a few hundred thousands in secondary school [\cite{edu}]. The government has been key in providing free education to everyone; unfortunately it has been unable to keep up with the programme of school construction and teacher training. Post-school education is still limited and very competitive. There are only two state and five private universities and 15 superior institutes [\cite{uni}]. The undergraduate degree courses last up to 4 years (for a {\it Licenciatura}). Only in the last couple of years have some Masters courses appeared, but as yet there is no accreditation for Doctoral studies.

\begin{figure}[h]
\centering
\includegraphics[width=2in]{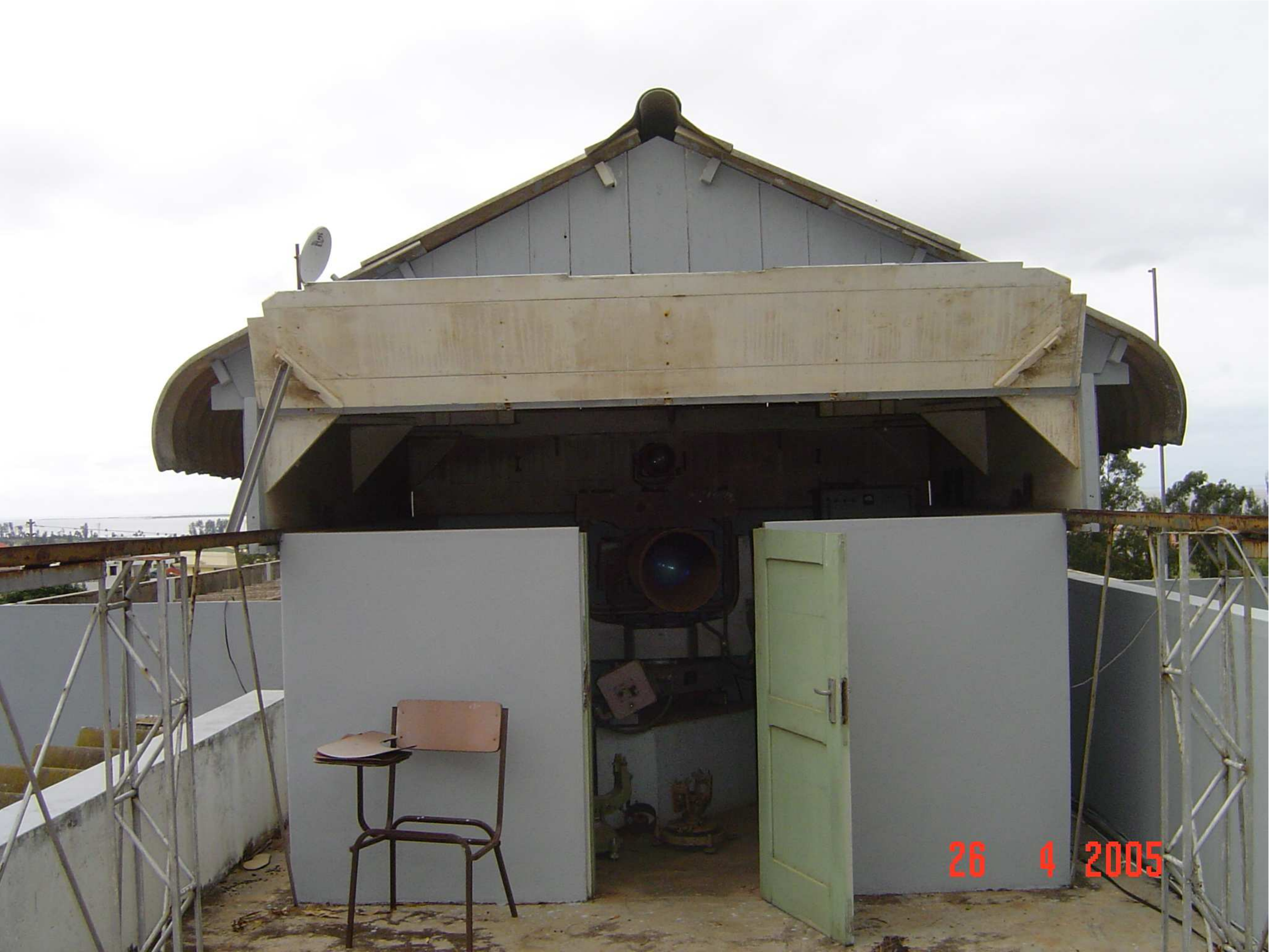}
\includegraphics[width=2in]{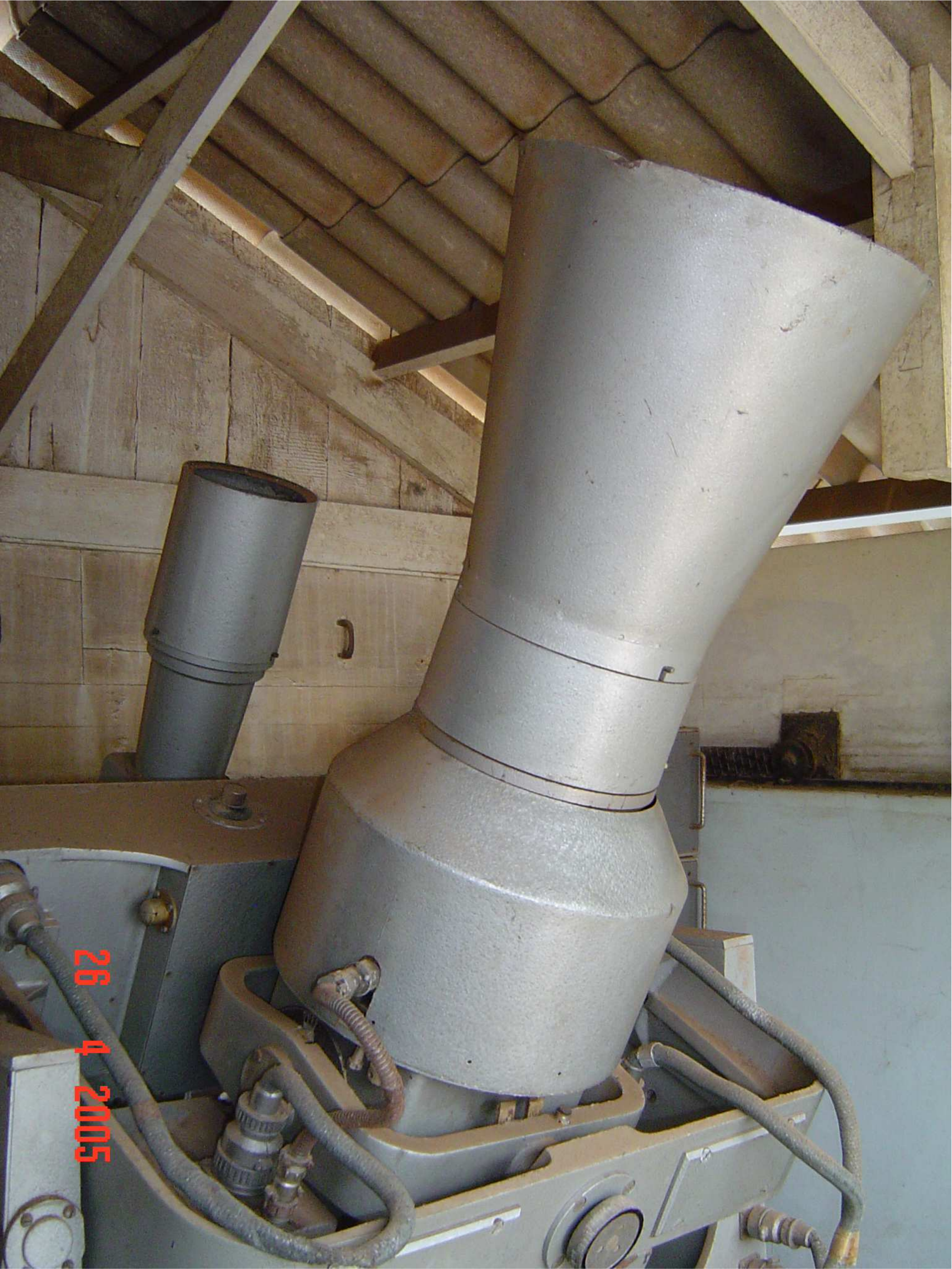}
\caption{Inactive telescope and dome enclosure at the top of the Mathematics Department at the Universidade Eduardo Mondlane}
\label{fig:2}
\end{figure}

The earliest recorded history of astronomical observations and interest came from the Portuguese Military in the early 1900's where officers were tasked to observe the solar eclipse and the night sky. In 1907 Frederico Oom was assigned with acquiring material and building an observatory in the then Louren\c{c}o Marques, now Maputo [\cite{hist}]. It is not clear what subsequently happened here, however, there is an inactive telescope on top of the Mathematics Department at the main university in Maputo, Universidade Eduardo Mondlane (UEM; Figure \ref{fig:2}). In 1997 an academic at UEM and some students tried to create conditions for astronomy to have a place at the university however, they were unsuccessful. In 2004-2005 another attempt by the N\'ucleao de Estudantes de F\'isica at UEM, in collaboration with the Portuguese Superior Institute, sparked more interest, but also it did not go very far.

\section{International Year of Astronomy 2009}
\begin{figure}[h]
\centering
\includegraphics[width=4.5in]{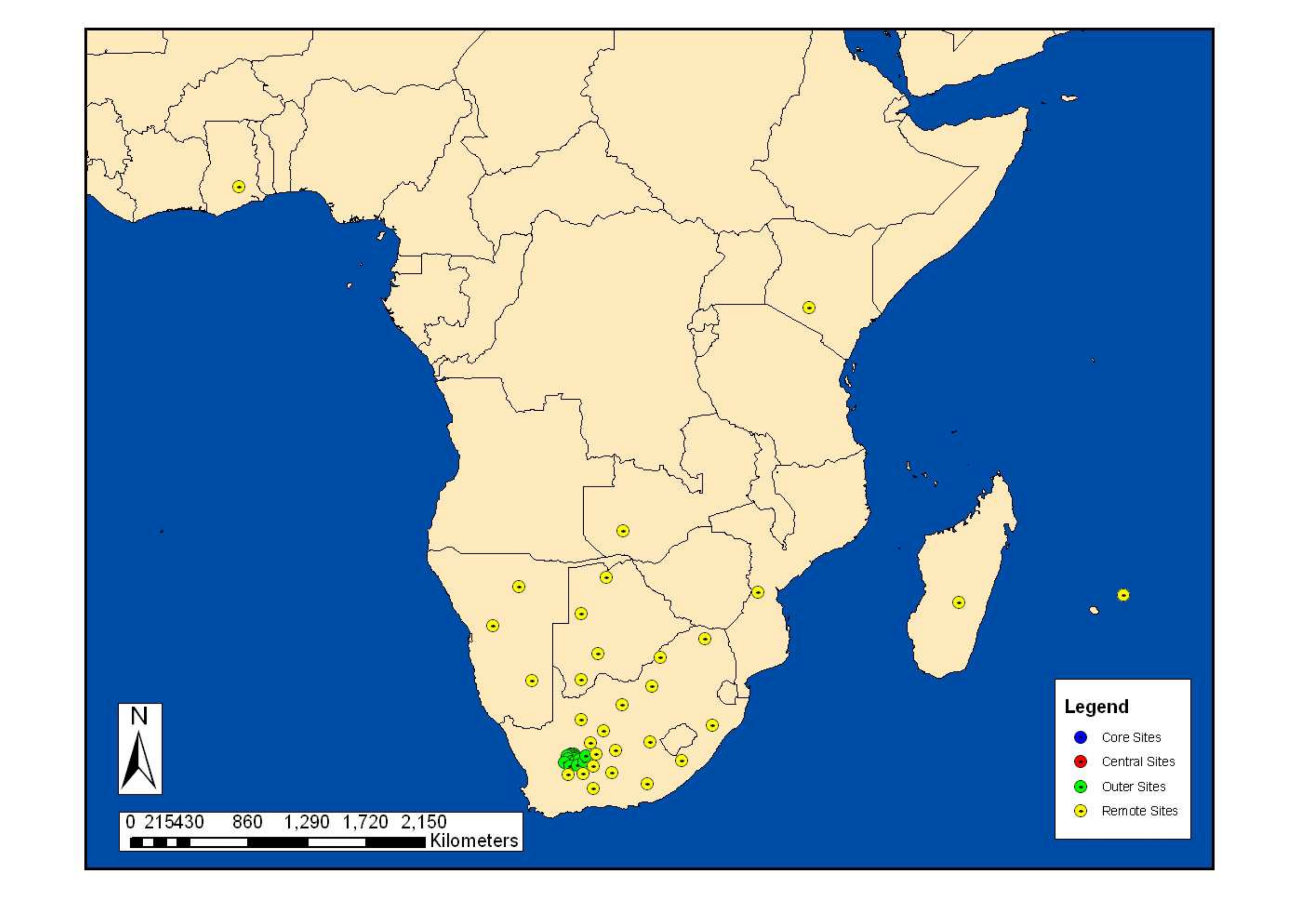}
\caption{Proposed configuration of the core site (green) for the Square Kilometre Array in South Africa's Northern Cape, with outstations (yellow) in Namibia, Botswana, Mozambique, Zambia, Kenya and Ghana, as well as on Madagascar and Mauritius. Source: http://www.ska.ac.za/media/visuals.shtml, accessed 2009 May 13.}
\label{fig:3}
\end{figure}

With the announcement that 2009 was to be the International Year of Astronomy (IYA) by UNESCO, one of the authors, CM Paulo, took the post of Single Point of Contact for Mozambique while an MSc student at the University of Western Cape, South Africa, funded by the South African Square Kilometre Array (SKA) project. This project has been funding several people from across Africa at MSc, PhD and Post-doctoral positions and due to the size of the project, Mozambique has been suggested to host an out-station (Figure \ref{fig:3}). Another author, F Nhanombe, has also been funded by this project.

Mozambique planned a series of activities for IYA, some of these drawn from the 2004-2005 attempt, mainly aimed at school level, some of which started pre-IYA. There is hope that these activities will branch out to other peer groups as funding is sought. It is important to mention that these activities have been carried out with little or no funding and the equipment used has been loaned by a local school and amateur astronomers.

\subsection{Activities}
The planned activities and/or goals involve:
\begin{itemize}
\item seminars and observations in local schools, by local and international speakers about astronomy during the months of March, June and October. The June activities have significant importance because these are planned for June 1st, Child Day in Mozambique.
\item exhibitions at the Ministry of Science and Technology; these will mainly be at various events organised by the ministry and through publications in the ministry's magazine.
\item coverage of the activities and efforts in the media.
\item gathering information about the amateur astronomy community and creating conditions for the establishment of an Amateur Astronomy Society.
\item creating conditions for the development of astronomy professionally in Mozambique.
\end{itemize}

As a starting place, pre-IYA, some activities have been accomplished. In June 2008, there were seminars by Dr. Peter Martinez, from the South African Ministry of Science and Technology, about how the International Astronomical Union can help establish astronomy and the possibilities of collaborations between the South African and Mozambican governments. In October 2008, during the World Space Week, the first attempt at an IYA activity was launched and the Portuguese School and Estrela Vermela School, which coincidentally means Red Star, were involved in demonstrations, singing and observations of the moon during the evening.

The first activity of 2009 took place at the Portuguese School on the evening of April 24. In attendance were children from the Portuguese, Estrela Vermelha and Italian schools with academics and students from the Department of Physics at UEM and dignitaries from the Italian Embassy. The evening involved a performance by students from the Portuguese School about our solar system and then observations of Saturn through a telescope.

\section{Future Plans}
Here we identify the obvious future plans, and at times ambitious ones:
\begin{itemize}
\item An introductory astronomy course will be introduced in the Physics Department at UEM for their 2nd year students in 2010. This will comprise of 2 hours of lectures a week for 16 weeks. The course will include subjects such as why astronomy is not astrology, why there are seasons and, on a much larger scale, cosmology.
\item Establish an Amateur Astronomical Society, hopefully run by the amateurs themselves.
\item Explore how Mozambique and the International Astronomical Union can co-operate, for example through Commission 46 ``Teaching Astronomy for Development".
\item Investigate further the history of astronomy in Mozambique, mainly pre-1900 and also the gap in the 20th century. We plan to contact the Lisbon Observatory who can provide some answers to this.
\item Secure long term, or at least steady, funding.
\item In terms of ambitious long term projects, introduce further courses at the university, study the potential to establish an astronomical observatory and integrate astronomy into society fully.
\end{itemize}

\section{Conclusions}
This paper has hopefully shown the enthusiasm of a young and dynamic team that has been working at different levels to introduce astronomy into Mozambican Society. Mozambique falls under phase 3, countries that have a non-existent astronomy community but show strong potential in the form of physics/mathematics research and outreach communities who are willing to drive the development of astronomy, as defined by the Developing Astronomy Globally cornerstone project for IYA, and is working towards improving this.

The authors see the IYA as an excellent opportunity to bring astronomy into Mozambican society and hopefully get the amateur community fully engaged. Planning is well under way, for 2010, to have an introductory astronomy course at UEM, which will hopefully get more students involved or interested in this subject area. The question will be, however, how long can this IYA momentum be carried beyond 2009 and will we continue to see positive effects in later years?

\acknowledgements 
The authors would like to thank Michael Bode, Phil James and Ant\'{o}nio Ribeiro for useful comments and discussions.

The attendance of V.A.R.M.R. at the IAUS260 was made possible through funding from the Royal Astronomical Society, Liverpool John Moores University Astrophysics Research Institute and the UK's Institute of Physics.

\end{document}